# Metastability limit for the nucleation of NaCl crystals in confinement

*Julie Desarnaud [a], Hannelore Derluyn [b], Jan Carmeliet [b], Daniel Bonn [a] and Noushine Shahidzadeh [a\*]*

[a] Van der Waals-Zeeman Institute, Institute of Physics, University of Amsterdam, Science Park 904, 1098 XH Amsterdam, The Netherlands

[b] ETH, Institut für Technologie in der Architektur, HIL E 46.3, Wolfgang-Pauli-Str. 15, 8093 Zürich Hönggerberg, Switzerland.

**Corresponding Author: n.shahidzadeh@uva.nl**

ABSTRACT-We study the spontaneous nucleation and growth of sodium chloride crystals induced by controlled evaporation in confined geometries (microcapillaries) spanning several orders of magnitude in volume. In all experiments, the nucleation happens reproducibly at a very high supersaturation S~1.6 and is independent of the size, shape and surface properties of the microcapillary. We show from classical nucleation theory that this is expected: S~1.6 corresponds to the point where nucleation first becomes observable on experimental time scales. A consequence of the high supersaturations reached at the onset of nucleation is the very rapid growth of a single skeletal (Hopper) crystal. Experiments on porous media reveal also the formation of Hopper crystals in the entrapped liquid pockets in the porous network and consequently underline the fact that sodium chloride can easily reach high supersaturations, in spite of what is commonly assumed for this salt.

**TOC GRAPHICS**

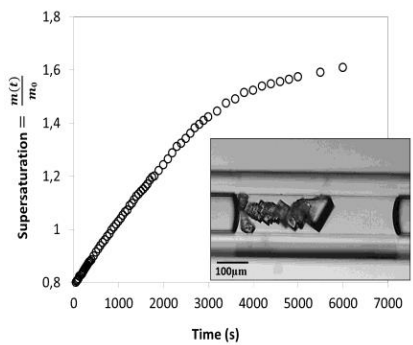

Sodium chloride is the most abundant salt on earth, and its crystallization is very important in many applications. Degradation and desertification of soils due to sodium chloride is a major physiological threat to ecosystems (1-3). Moreover, its crystallization in confined conditions is known to be not only one of the major causes of physical weathering and disintegration of rocks, stones and building materials (4-10) but also constitutes a problem for oil well productivity and $CO_2$ storage due to increased impermeability of deep soil layers and rocks (11,12). On the other hand sea salt aerosols are one of the most abundant primary inorganic aerosols and their kinetics of deliquescence/crystallization provide important insights to the alteration of the particle aerodynamic properties and their cloud-droplet nucleation efficiency (13). For most if not all of the above mentioned applications, the precise conditions under which NaCl crystals nucleate and growth from solution are very important but largely unknown.

Although recent simulations give us some more insight into the atomistic pathways of the nucleation of NaCl crystals (14,15), the current consensus appears to be that no high supersaturations can be reached for this salt, and that the pore size has a profound influence on the concentration at which the nucleation is first observed (4,16-20), in spite of the fact that typical pore sizes are orders of magnitude larger than the size of a critical nucleus.

Here, we report experiments on the primary nucleation and growth of sodium chloride crystals by controlled evaporation for different degrees of confinement in microcapillaries in which the salt solution is trapped by capillary forces. We use confined geometries consisting of glass microcapillaries of different sizes (from 20 to 2000 μm) with different surface chemistries: hydrophobic (silanised) and hydrophilic (cleaned) (21). The geometry of the microcapillaries is also changed; besides cylindrical we also use rectangular shapes since in real porous media liquid can be trapped within corners of the porous network (21-24). We study situations with slow evaporation (and consequently no large salt concentration gradients) by choosing the initial concentration in such a way that the surface tension and the contact angle of the salt solution was high enough to avoid the formation of wetting films..

We start out with aqueous solutions of known initial concentration $m_i$=4.9 mol.kg$^{-1}$ of NaCl (Sigma-Aldrich purity >99.9%) in Millipore water; the saturation concentration being $m_0$=6.15 mol.kg$^{-1}$; the relative supersaturation $S$ then defined as $m/m_0$ . The crystallization is induced by evaporation under isothermal conditions : a known volume V$_0$ ( ranging from 10$^{-3}$ to 5 μl) of the solution is introduced into the capillary and placed into a miniature climatic chamber under a microscope. By fixing the relative humidity of the ambient air in the climatic chamber (6), the evaporation rate of the solution is controlled. The volume change during the evaporation of the solutions inside the microcapillaries is subsequently followed by recording the displacement of the two menisci while simultaneously visualizing the onset of crystal growth in the solution directly with an optical microscope coupled to a CCD camera.

Fig. 1 shows the supersaturation at the onset of spontaneous crystal growth for different sizes, shapes and surface chemistry of the capillaries and for different evaporation rates.

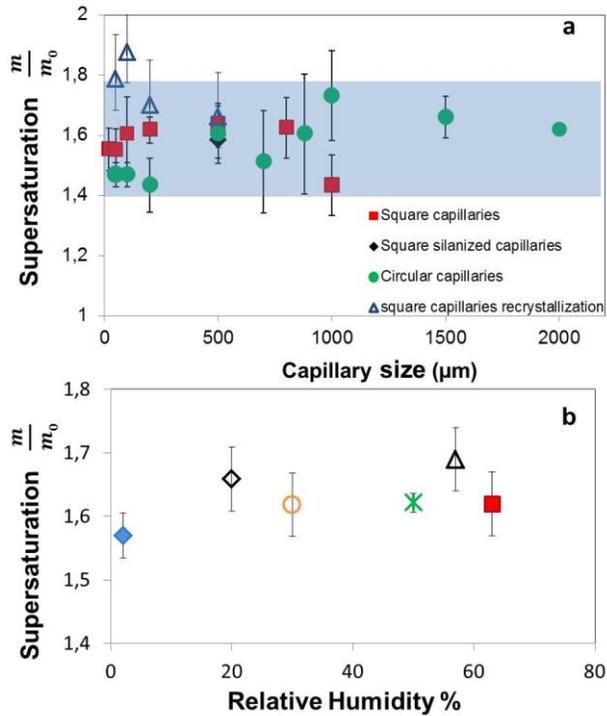

**Figure 1**. Supersaturation (S = $m/m_0$) of NaCl solutions reached by evaporation at the onset of nucleation and growth ($m$ and $m_0$ are molalities mol.kg$^{-1}$ at the onset of crystallization and at equilibrium respectively) :(a) data for microcapillaries of different sizes, geometries (square and circular) and wetting properties (cleaned and silanized) at RH~52±2%; (b) data at different relative humidities for square microcapilly 200 μm. Each point is an average of more than 8 experiments; the error bar indicating the spread in the observed values.

From more than 100 experiments, a limit of supersolubility, of S~1.6 ± 0.2 is found. This is very high compared to the limit of metastability reported in the literature for sodium chloride in cooling experiments, S=1.01 (25,26). To assess a possible effect of impurities on the concentration for which nucleation is first observed, recrystallization experiments were conducted by performing repeated cycles of complete deliquescence (dissolution by water vapour) of the salt crystals followed by drying, a procedure that is known to efficiently expel impurities (6,27). These results show again that the concentration reached at the onset of nucleation and crystal growth is not affected to within the experimental uncertainty (Fig 1(a)).

One difference between evaporation and cooling experiments is that, when the ion transport in the liquid phase is slower than the evaporation rate, there may be concentration gradients in the salt solution. This leads to a higher concentration of ions close to the meniscus, where the evaporation takes place. To see whether our supersaturation is well defined in terms of the NaCl concentration, we determine the Peclet number (Pe), which is a measure of the heterogeneity of the ion distribution. Pe is defined as the ratio between the convective and the diffusive transport of ions in the solution and can be calculated from the (time-dependent) NaCl diffusion coefficient and the characteristic time of the displacement of the meniscus (22):

$Pe \approx t_{diff}/t_s$ (1)

which in turn are given by $t_{diff} \approx z_0^2/D_{NaCl}(t)$ and $t_s \approx z_0/(dz/dt)$ with $z_0$ the initial meniscus position, z the position at time t, and $D_{NaCl}(t)$ the diffusion coefficient at time t, which depends on the concentration and viscosity of the solution (28).

We find that, Pe is of the order of unity at first, but reaches values on the order of $10^{-2}$ to $10^{-3}$ at the onset of nucleation. The latter underlines a homogeneous distribution of ions in the solution and is mainly due to the fact that the evaporation slows down, as shown in Fig. 3.

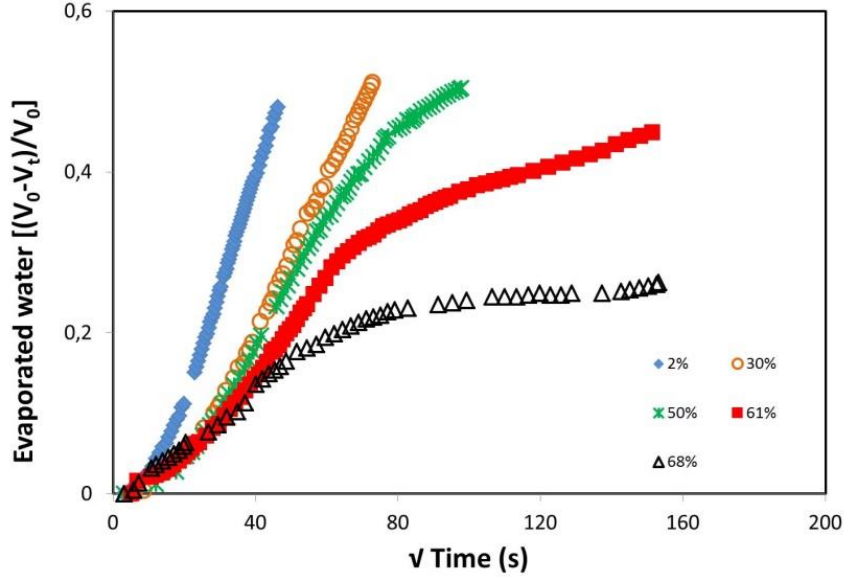

**Figure 3**.(a) Normalized evaporated water volume as a function of √time) at different relative humidities till spontaneous crystal growth is observed ($m_i$= 4.9 mol.kg$^{-1}$, d=200 µm square capillaries).

When the evaporation rate e is limited by diffusive vapor transport through the gas phase, it is given by (22,29):

$$e \approx \rho_g D \frac{(c_i - c_\infty)}{\delta} \qquad (2)$$

with $\rho_g$ the vapor density, $D$ the diffusion coefficient of water vapor through the gas, $c_\infty$ the controlled water vapour concentration of the climatic chamber and $c_i$ the water vapor concentration just above the menisci; δ is the characteristic distance over which diffusion takes place (22,29). The drying rate is consequently controlled by δ and ($c_i$-$c_\infty$) ; during the evaporation($c_i$-$c_\infty$) decreases because the saturated vapor concentration decreases when the salt solution becomes more concentrated (($c_i$/ $c_{pure\ water}$= 1-0.24 S (30)); on the other hand δ is roughly the distance between the meniscus and the outlet of the capillary, which increases: both effects then lead to a decrease in evaporation rate with time and consequently a simple diffusive $\sqrt{t}$ scaling is note expected (29,31).

The necessary time to reach the concentration at which the spontaneous nucleation and growth is observed depends on the relative humidity and varies over more than an order of magnitude. However, it is interesting to note that, independently of the time necessary to reach it, the crystallization is only observed when the supersaturation of 1.6 is achieved. As a consequence, for humidities higher than 61%, corresponding to the equilibrium water vapor concentration above a solution close to our limit of supersolubility (S~1.6), the evaporation rate goes to zero as soon as $c_i = c_\infty$ (see e.g. the black symbols in Figure 3). Here, in spite of the fact that the solution is supersaturated, nucleation is not observed because the supersaturation 1.6 is not reached. The metastable supersaturated solution can remain for days without any spontaneous crystallization until a perturbation is introduced to the system (i.e. rapid decrease of the temperature or water vapour concentration in the air) triggers the nucleation. This series of experiments independently provides another value for the onset of nucleation: S = 1.60 ± 0.07 (Fig. 1b), in agreement with the findings above.

Since the nucleation appears to be homogeneous, Classical nucleation theory (CNT) can be used to predict the rate of crystal nucleation for sodium chloride as a function of the supersaturation. According to CNT, the rate of nucleation per unit volume can be calculated as the product of an exponential factor and a kinetic prefactor (32,33):

$$J = \kappa \exp\left(-\frac{\Delta G^*}{kT}\right) \qquad (3)$$

In the exponential factor, $\Delta G^*$ is the free energy cost of creating a critical nucleus and $kT$ the thermal energy. The total Gibbs free–energy cost to form a spherical crystallite has a bulk and a surface term and can be expressed as:

$$\Delta G = \frac{4}{3}\pi R^3 \rho_s \Delta\mu + 4\pi R^2 \gamma \qquad (4)$$

With ρs is the number density of solid, Δμ the difference in chemical potential of the solid and liquid and $\gamma$ is the interfacial tension of the NaCl crystal with the solution ($\gamma_{lc}$~ 0.08 N.m$^{-1}$ (32,34)). Here, the difference in chemical potential of the solid and liquid, can be written in terms of the mean ionic activity of the solute a± (25), as :

$$\frac{\mu_l - \mu_c}{RT} = \nu \ln\left(\frac{a_\pm}{a_{0\pm}}\right) = \nu \ln\left(\frac{\gamma_\pm}{\gamma_{\pm 0}}\frac{m}{m_0}\right) \qquad (5)$$

where $m$ and $m_0$ are the molalities at crystallisation and equilibrium (mol.kg$^{-1}$) and γ± is the corresponding mean ionic activity coefficient and ν is the sum of ions (2 for NaCl)

The energy of the critical nucleaus ΔG* is then:

$$\Delta G^* = \frac{4\pi}{3}\gamma R_c^{*2} \quad \text{with } R_c^* = \frac{2\gamma}{\nu\, kT \ln\left(\frac{\gamma_\pm}{\gamma_{\pm 0}}\frac{c}{c_0}\right)} \qquad (6)$$

The kinetic prefactor κ, which relates the efficiency with which collisions between supernatant ions and the crystal interface produce crystal growth is determined from κ=ρjZ, where ρ is the number density of molecules in the liquid phase, Z the Zelodovic factor : $Z \approx 1/(n^*)^{2/3}$ with n* the excess number of molecules in the critical nucleus and $j$ the rate at which molecules attach to the nucleus causing its growth. $j$ is approximated as $j \sim \rho D R^*$ with D the diffusion constant of the molecules and R$_c^*$ the radius of the critical nucleus (33).

The rate of nucleation $J$ (m$^{-3}$s$^{-1}$) is plotted as a function of the supersaturation in Figure 4. It follows that, below S~1.6 the nucleation rate is extremely small and conversely, very large above it. At S~1.6 the rate of the nucleation of sodium chloride is found to be 0.004 m$^{-3}$s$^{-1}$, which roughly corresponds to one nucleus in a typical volume considered here within our experimental time window (typically 5000 s). Moreover, the variation of the experimental parameters here (volume $V$, relative humidity etc.) do not significantly change the value of S~1.6 for which the nucleation becomes observable. Simply said, the nucleation rate depends so steeply on the supersaturation that all other parameters are irrelevant, in excellent agreement with all our observations.

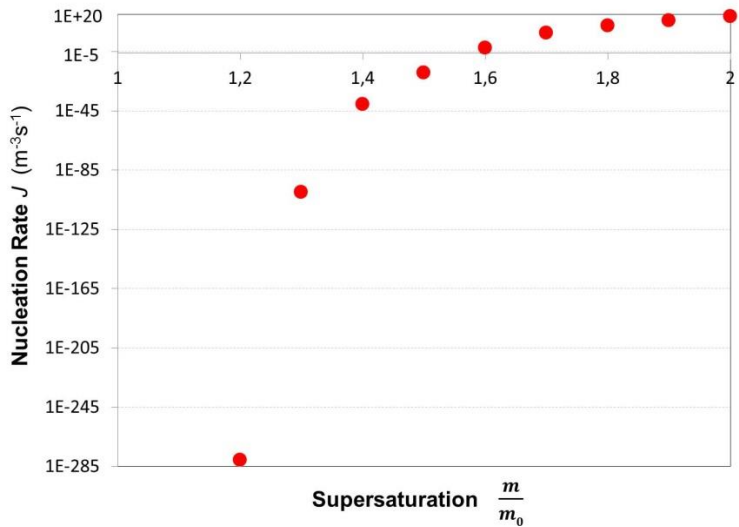

**Figure 4.** Nucleation rate J ($m^{-3}s^{-1}$) as a function of supersaturation $S=m/m_0$ of the solution.

Another very interesting observation is that at the onset of crystallization a single nucleus is observed to be growing very rapidly and with a peculiar shape: a Hopper (skeletal) crystal (Figure 5). The growth of a single crystal is likely to be an effect of the confinement: in a small size system, the growth of the critical nucleus will lead to a very rapid decrease of the local supersaturation and consequently favors the formation of only a single crystal (35). The growth of the Hopper crystals in these experiments happens at a speed that can be up to ten times that of the growth of a regular cubic crystal under the same conditions (6). The rapid growth of the Hopper crystal is due to the high supersaturations for which secondary nucleation may occur at the corners of the growing primary nuclei, due to the disparity of growth rates between the crystal edges and the crystal faces (36,37). In our experiment, this shows up as the rapid formation of a chain-like structure of crystals that are all interconnected (Figure 5).

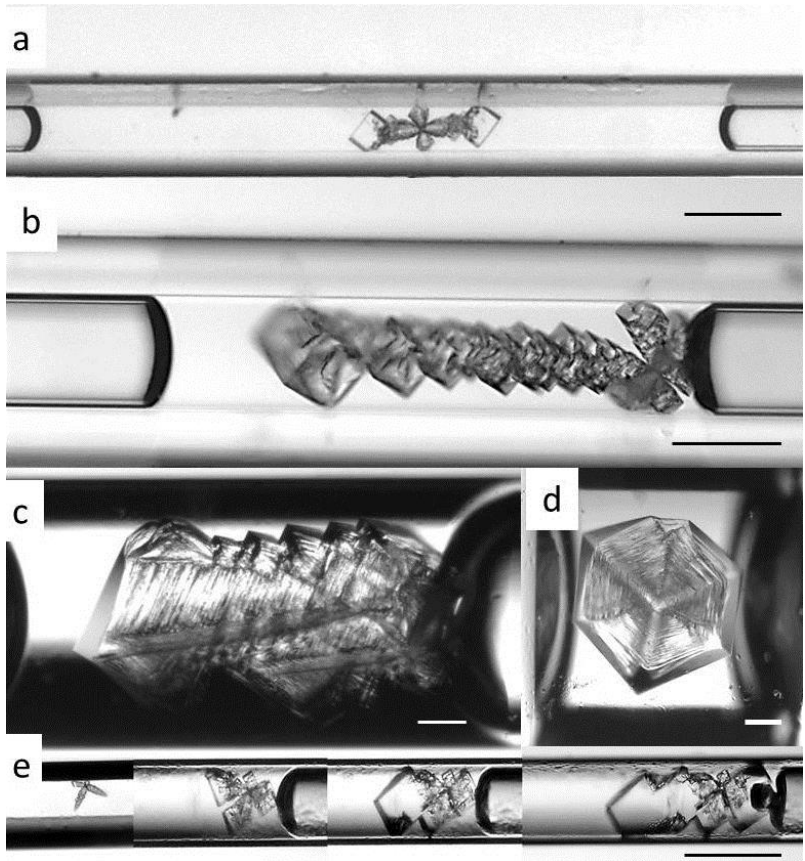

**Figure 5**. Spontaneous growth of Hopper crystals at supersaturation S~1.6, in capillaries: (a) 50μm (b) 100μm (c) 500μm (d) 1000μm. (e) Evolution of the growth of a Hopper crystal in the first 10 minutes (Scale=100μm).

As the capillaries are often used as a model system for crystallization in pores, it is likely that such fast growth dynamics may provoke damage in porous materials; Indeed, the crystallization pressure that is responsible for the damage is known to depend strongly on the supersaturations reached (38), the speed of growth (40) and the size of the crystal (39).

The morphology of the Hopper crystals is observed to change during the growth process, due to the decrease of the ion concentration in the solution accompanying the formation of the crystals. At the late stages of growth large regular cubic crystals start to form from the extremities of the Hopper crystal (figure 5e).

To see whether our conclusions from the capillaries also apply to real porous media, we compare the crystal morphologies in microcapillaries at the late stages of drying with the NaCl crystals formed in experiments on porous sandstone (Prague sandstone with average pore diameter d~30μm and porosity ~29% (5)). To be able to compare the two situations (capillaries and sandstone), the formation of entrapped liquid pockets in the porous network was facilitated by treating the surface of the stone with a water repellent product (silanes). The treatment slows down the evaporation, prevents salt from crystallizing at the exterior of the stone (23) and facilitates the formation of liquid pockets. After saturation of the stone with the initial salt solution ($m_i$=4.9 mol.kg$^{-1}$), the latter is dried under the same environmental conditions as the capillaries. Subsequently, the sample is fractured and the salt crystal morphology inside the stone is investigated by Scanning Electron Microscopy (SEM).

Figure 6 shows the remarkable similarities between the crystal structures formed in the stone and in the capillaries : in the stone also Hopper crystals have formed. Observations on several samples show that these crystals shape are plentiful in some regions of the stone, and almost absent from others, suggesting that very concentrated residual fluid pockets had formed during the evaporation. Both these observations indicate for the first time that the liquid pockets in the stone behave similarly to the microcapillaries where high supersaturations were reached before crystallization sets in.

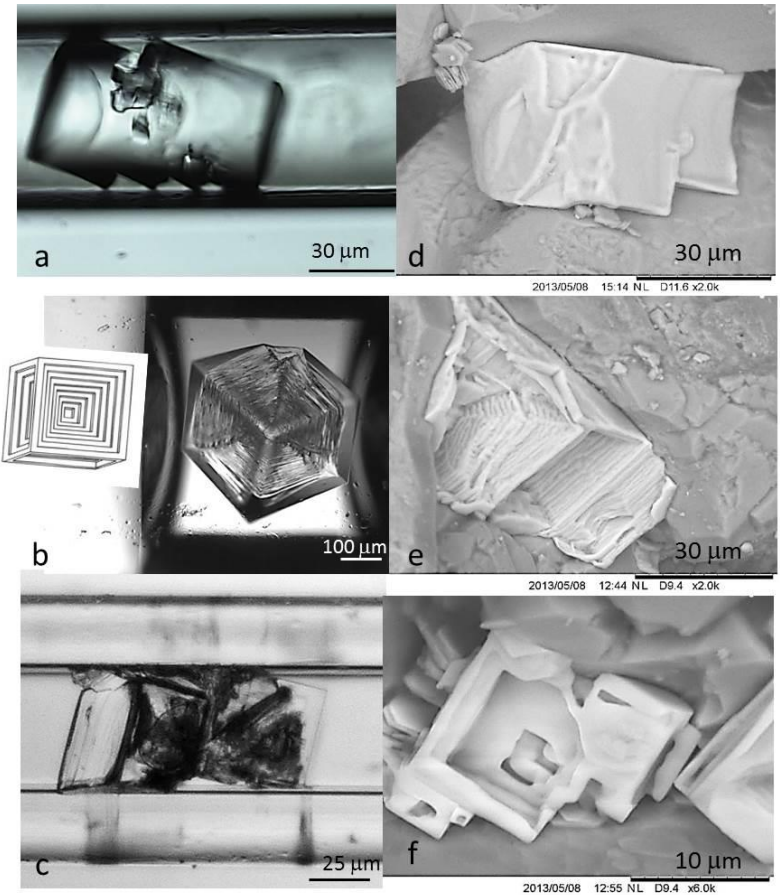

**Figure 6**. Comparison between Nacl crystals formed in capillaries after reaching t our limit of supersolubility S1.6 (a-c) and those obtained in the sandstone (d-f).

In sum, we have demonstrated by controlled evaporation experiments of sodium chloride solutions, that the supersaturation achieved at the onset of spontaneous primary nucleation and growth is around 1.6 and remains independent of the size, shape and surface properties of the microcapillary. These results are consistent with expectations from classical nucleation theory. The supersolubility limit obtained here clearly shows that, contrary to what is commonly assumed for this salt (4,16,17,19,41), high concentrations can be reached before spontaneous crystal growth. This in turn leads to the formation of a Hopper crystal, which we also detected in analogous experiments conducted on real sandstone. Our findings therefore have far-reaching implications for the widespread consequences of salt crystallization (1-12, 42), since the salt weathering of rocks, stones and monuments is related to the crystallization pressure which is directly dictated by the supersaturation and the importance of which is expected to increase in the future due to global climate change (43).

**REFERENCES**

(1) Rengasamy, P.; Olsson, K. A. Sodicity and soil structure. *Aust. J. Soil. Res*. **1991,** 29, 935-952.

(2) Wong, V. N., Dalal, R. C., Greene, R. S. Salinity and sodicity effects on respiration and biomass of soil. *Biol. Fert. Soils*. **2008,** 44**,** 943-953.

(3) Cook, R. U.; Smalley, I. J. Salt weathering in deserts. *Nature*. **1968**, 220, 1226-1227.

(4) Steiger, M. Growth in porous materials – I: The crystallization pressure of large crystals. *J. Cryst. Growth*. **2005**, 282, 455-469.

(5) Shahidzadeh-Bonn, N.; Desarnaud, J.; Bertrand, F.; Chateau, X.; Bonn, D. Damage in porous media due to salt crystallization. *Phys. Rev. E*. 2010, 81, 066110 .

(6) Shahidzadeh, N.; Desarnaud, J. Damage in porous media: role of the kinetics of salt (re)crystallization. *Eur. Phys. J. Appl. Phys*.**2012, 60**, 24205.

(7) Lubelli, B.; van Hess, R. P. J.; Huinink, H. P.; Groot, C. J. W. P. Irreversible dilation of NaCl contamined lime-cement mortoar dur to crystallization cycles. *Cement Concrete Res*. **2006**, 36, 678-687.

(8) Espinosa-Marzal, R.; Scherer, G.W. Impact of in-pore salt crystallization on transport properties. *Environ Earth Sci*. **2013**, 69, 2657-2669.

(9) Schiro, M.; Ruiz-Agudo, E.; Rodriguez-Navarro, C. Damage Mechanisms of Porous Materials due to In-Pore Salt Crystallization. *Phys Rev. let.* **2012,** 109, 265503.

(10) Veran-Tissoires, S.; Marcoux, M.; Prat, M. *Phys.Rev. Lett*. **2012**, 108, 054502.

(11) Muller, N., Qi, R., Mackie, E., Pruess, K., Blunt, M. J. $CO_2$ injection impairment due to halite precipitation. *Energy Proc.*2009, **1**, 3507-3514

(12) Peysson, Y. Permeability alteration induced by drying of brines in porous media. *Eur. Phys. J. Appl. Phys*. 2012 **60**, 24206

(13) Li, X.; Dhrubajyoti, D.; Hyo-Jin, E.; Hye Kyeong, K.; Chul-Un, R. Deliquescence and efflorescence behavior of individual NaCl and KCl mixture aerosol particles. *Atmosphere Environment*, **2014**, 82, 36-43

(14) Zahn, D. Atomistic mechanism of NaCl nucleation from an aqueous solution. *Phys.Rev.Lett.***2004,** 92, 040801.

(15) Mucha, M.; Jungwirth, P. Salt crystallization from an Evaporating aqueous solution by molecular dynamics simulations. *J. Phys. Chem. B*. **2003**, 107,33, 8271.

(16) Emmanuel, S.; Berkowitz, B. Effect of pores size controlled solubility on reactive transport in heterogeneous rock. *Geophys. Res. Lett*. **2010,** 34, L06404.

(17) Putnis, A.; Mauthe, G. The effect of pore size on cementation in porous rocks. *Geofluids*. **2001**,1, 37-41 (2001).

(18) Flatt, R. J. Salt damage in porous materials: how high supersaturations are generated. *J. Cryst. Growth*. **2002**, 242, 435-454.

(19) Bouzid, M.; Mercury, L.; Lassin, A.; Matray, J. M. Salt precipitation and trapped liquid cavitation in micrometric capillary tubes. *J. Colloid and Interf. Sci*. **2011**, 360,768-776.

(20) Sekine, K.; Okamato, A.; Hayashi, K. In situ observation of the crystallization pressure induced by halite crystal growth in a microfluidic channel. *American Mineralogist*. **2011**,96, 1012-1019.

(21) Shahidzadeh-Bonn, N.; Rafai, S.; Bonn, D.; Wegdam, G. Salt crystallization during evaporation: Impact of interfacial properties. *Langmuir*. **2008**, 24, 8599-8605.

(22) Camassel, B.; Sghaier, N.; Prat, M.; Nasrallah, S. B. Evaporation in a capillary tube of square cross-section: application to ion transport. *Chem. Eng. Sci.* **2005**, 60, 815-826.

(23) Wong, H.; Morris, S.; Radke, C. J. Three dimensional menisci in polygonal capillaries. *J. Colloid. Interf. Sci.* **1991,** 48, 317.

(24) Chauvet, F.; Duru, P.; Geoffroy, S.; Prat, M. The three periods of drying of a single square capillary tube. *Phys. Rev. Lett.* **2009,** 18, 124502.

(25) Mullin, J. W. *Crystallization* (Butterwords-Heinemann, 4th edition) **2001**.

(26) Chlanese, A.; DI Cave, S.; Matzarotta. Solubility and Metastable Zone Width of Sodium Chloride in Water-Diethylene Glycol Mixtures. *J. Chem. Eng. Data,* **1986** 31, 329-332.

(27) Desarnaud, J.; Shahidzadeh-Bonn, N. Salt crystal purification by deliquescence/crystallization cycling. *Euro. Phys. Lett.* **2011**, 95, 48002.

(28) Kestin, J.; Khalifa, E.; Correia, R. J. Tables of the dynamics and kinematic viscosity of aqueous NaCl solutions in the temperature range 20-150ºC and the pressure range 0.1-35 MPa. *J. Phys. Chem. Ref. Data.***1981**, 10, 71-87.

**(29)** Bird, R. B., Stuwart, W. E., Lightfoot, E. N. *Transport phenomena* (Wiley, New York) **1960**

(30) Robinson, R. A. The vapour pressures of solutions of Potassium Chloride and Sodium Chloride. *Trans. Roy. Soc. New Zealand.* **1945**, *75*, 203-217.

**(31)** Hołyst, R .; Litniewski, M .; Jakubczyk, D .; Kolwas, K .; Kolwas, M .; Kowalski, K.; Migacz, S .; Palesa, S.; Zientara, M. Evaporation of freely suspended single droplets: experimental, theoretical and computational simulations, *Rep. Prog. Phys.* **2013,** 76

(32) Valeriani, C.; Sanz, E.; Frenkel, D. Rate of homogeneous nucleation in molten NaCl. *J.Chem.Phys.* **2005,** 122, 194501

(33) Sear, R, Nucleation: theory and applications to protein solutions and colloidal suspensions. *J. Phys. Condens. Matter.* **2007**, 19, 033101

(34) Han-Soo, N.; Stephen, A.; Mayerson A.S. Cluster formation in highly supersaturated solution droplets. *J. Cryst. Growth.* **1994**, 139, 104-112

(35) Grossier, R.; Veesler, S.; Reaching one single and stable critical cluster through finite-sized systems. *J.Cryst.Growth and Design*. **2009** 9, 1017-1922

(36) Kuroda, T.; Irisawa, T.; Ookawa, A. Growth of a polyhedral crystal from solution and its morphological stability. *J. Cryst. Growth.* **1972,** 42, 41-46.

(37) Sunagawa, I. Growth and morphology of crystals. *Forma.* **1999, 14**, 147-166.

(38) Correns, C. W. Growth and dissolution of crystals under linear pressure. *Discussions of the Faraday Soc*. **1949**, 5, 267–271.

(39) Steiger, M. Crystal growth in porous materials ; Influence of crystal size on the crystallization pressure, *J. Cryst. Growth*. **2005**, 282, 470-481.

(40) Chatterij, S. Aspects of generation of destructive crystal growth pressure, *J.Cryst. Growth.* **2005**, 277**,** 566–577.

(41) Sghaier, N.; Prat, M.; Ben Nasrallah, S. On ions transport during drying in a porous medium. *Trans.Porous Med*. **2007**, 67, 243-274.

(42) Kiyohara, K.; Takushi Sugino, A.; Asaka, K. Electrolytes in porous electrodes: Effects of the pore size and the dielectric constant of the medium. *J. of Chem.Phys.* **2010,** 132, 144705.

(43) Climate change and world heritage, Editor: Agustin Colette, UNESCO World Heritage Centre, **2007**